# A Distributed File System for a Wide-Area High Performance Computing Infrastructure

Edward Walker
*University of Texas at Austin*

## Abstract

We describe our work in implementing a wide-area distributed file system for the NSF TeraGrid. The system, called XUFS, allows private distributed name spaces to be created for transparent access to personal files across over 9000 computer nodes. XUFS builds on many principles from prior distributed file systems research, but extends key design goals to support the workflow of computational science researchers. Specifically, XUFS supports file access from the desktop to the wide-area network seamlessly, survives transient disconnected operations robustly, and demonstrates comparable or better throughput than some current high performance file systems on the wide-area network.

## 1. Introduction

How do computational scientists access files in a widely-distributed computing environment? And what needs to be provided to enable them to access files easily and be productive in these environments? We provide one data point from the NSF TeraGrid towards answering the first question and we propose a system that we believe provides a simple answer to the second question.

The TeraGrid is a multi-year, multi-million dollar, National Science Foundation (NSF) project to deploy the world's largest distributed infrastructure for open scientific research [1]. The project currently links over 9000 compute nodes, contributed from 19 systems, located at nine geographically distributed sites, through a dedicated 30 Gbps wide-area network (WAN). In aggregate, the project provides over 100 TFLOP of compute power and four PB of disk capacity for computational science research. Future system acquisitions for the project are expected to expand the compute capability of the infrastructure into the PFLOP range.

In 2005, TeraGrid conducted a survey amongst 412 of its users. One of the results from the survey highlighted the secure copy command, SCP, as the most important data management tool currently being used [2]. To address this issue, the GPFS-WAN distributed parallel file system [3] went into production that same year at three sites on the TeraGrid: the San Diego Supercomputing Center (SDSC), the National Center for Supercomputing Applications (NCSA) and Argonne National Laboratory (ANL). Users and jobs running on any compute node on systems at these three sites could now access their files from a common /gpfs-wan/ directory cross-mounted at all three sites. GPFS-WAN was selected over other wide-area distributed file systems, like OpenAFS [4] and NFSv4 [5], because of the high speed parallel file access capabilities provided by the underlying GPFS technology [6][7]. Some of the largest supercomputers in the world currently use GPFS as their primary work file system.

However, a number of issues persist with this solution. First, files (data and code), residing on the user's personal workstation, still needs to be manually copied to at least one of the TeraGrid sites, causing multiple versions of these files to be created. Second, not all sites use GPFS as their primary work file system, precluding these sites from participating in the GPFS-WAN solution. Other parallel file system technologies in use include Lustre [8] and IBRIX [9].

In this paper, we describe a distributed file system we have implemented to address these concerns, and in the process extend some of the traditional design goals assumed by many past distributed file systems projects. Many techniques proposed by past projects are valid today, and we have therefore incorporated these into our design. But there is an opportunity to reexamine some of these past design goals, incorporating the current reality of personal mobility, hardware commoditization and infrastructure advancements, into a solution that not only builds on prior art, but introduces new, more productive, usage patterns.

Today, the commoditization of disk and personal computers has resulted in many scientists owning personal systems with disk capacity approaching hundreds of GB. These computational scientists have the ability to routinely run simulations and store/analyze data on a laptop computer or a desktop workstation. Only the largest compute runs are scheduled on high-end systems at the NSF supercomputing centers.

Also, wide-area network bandwidth is no longer as scarce a resource as it once was in the computational science community. In addition to the TeraGrid network, other national projects like NLR [10] and Abilene-Internet2 [11] are providing multi-Gbps wide-area networks, with the goal of 100 Mbps access to every desktop. Furthermore, a lot of this available bandwidth is not currently fully utilized [12], so there is an opportunity for software systems to better utilize this spare capacity, and provide for a richer computing experience.

In light of these conditions, this paper reconsiders the following often cited distributed file system design goals [14][15][16][17][30]:

**a. Location transparency**: All past distributed file systems projects assumed this goal. Although being able to walk down a corridor to physically access the same files from a different workstation may still be important, a scientist is equally mobile by physically carrying his (or hers) personal system in the form of a laptop, or a memory stick, into a different network. Thus, the ability to extend a users personal work-space from a laptop, or a foreign desktop with an attached memory stick, is equally important.

**b. Centralized file servers**: Many systems, such as AFS [14], Decorum [15] and Coda [16], make a strict distinction between clients and servers. Some treat clients primarily as memory caches, like Sprite [17]. However, personal systems now have large disk capacity, and this is changing the role of the "client". Files residing on multiple, distributed, personal systems need to be exported to a few central systems for access by users and their applications. In other words, these large central systems are now the "clients" for files residing on the user's personal workstation.

**c. Minimize network load**: Systems like Decorum and Sprite strive to minimize network load as one of their fundamental design goals. Although this is still an important design goal, it is no longer as critical as it once was. Cost-benefit tradeoffs may be made to determine if the increased user productivity justifies the cost of the extra network bandwidth utilized.

**d. File sharing between users**: Previous studies [19][20] have shown that file sharing between users on a file system is not common. However, many distributed file systems still consider it a major benefit to provide the ability to share files between users [21]. Recently, many other non-file system mechanisms have been developed to facilitate wide-area data sharing: e.g. web portals, peer-to-peer systems, etc. Therefore, developing a system to enable *private*, transparent, access to personal files across remote sites not only addresses the most common use case, but also simplifies the design space considerably.

For the remainder of this paper, we describe the XUFS distributed file system. XUFS is a distributed file system that takes the assumption of personal mobility, with significant local disk resource and access to high-bandwidth research networks, as important considerations in its design criteria. Section 2 further expounds on the system requirements, including motivating empirical observations made on the TeraGrid. Section 3 describes the component architecture, the cache coherency protocol, the recovery mechanism, an example access client, and the security framework used in XUFS. Section 4 looks at some comparison benchmarks between XUFS and GPFS-WAN, and section 5 examines related work. Section 6 concludes this paper.

## 2. Requirements Overview

### 2.1. Computational Science Workflow

From our interviews with many scientists, the computational science workflow shows many commonalities. This workflow usually involves 1) developing and testing simulation code on a local resource (personal workstation or departmental cluster), 2) copying the source code to a supercomputing site (where it needs to be ported), 3) copying input data into the site's parallel file system scratch space, 4) running the simulation on the supercomputer (where raw output data may be generated in the same scratch space), 5) analyzing the data in the scratch space, 6) copying the analysis results back to the local resource, and 7) moving the raw output to an archival facility such as a tape store.

In this scenario, files move from the local disk to a remote disk (where they reside for the duration of the job run, and final analysis), with some new files moving back to the local disk, and the remainder of the new files moving to some other space (like a tape archive).

In this usage pattern, a traditional distributed file system will not suffice. Some simulations generate very large raw output files, and these files are never destined for the user's local resource. However, users still require transparent access to these large data files from their personal work-space in order to perform result analysis [24].

### 2.2. File-Sharing in Computational Science

Previous studies have shown that file-sharing between users is not common [19][20]. However in computational science this appears to be even less prevalent. We looked at the directory permissions for the 1,964 TeraGrid users in the parallel file system scratch space on the TeraGrid cluster at the Texas Advanced Computing Center (TACC).

We looked at this cluster because of the site's policy of setting the default `umask` for every user to "077". Thus a directory created by the user will only have user read-write permission enabled by default. Group read-write permission needs to be explicitly given by the user to permit file sharing. Other sites have a more lenient `umask` policy; hence no conclusion can be derived by simply looking at the directory permissions at those sites. However, since the directories we examined at TACC belonged to TeraGrid-wide users, with many having the same accounts on the other systems on the TeraGrid, our conclusion represents a valid data point for the entire project as a whole.

Of the 1,964 directories examined at the time of this paper preparation, only *one* user explicitly enabled group read-write permission. This of course does not preclude file sharing in the future through a different work space, web site or central database. However, we conclude that at least for the duration covering the period of input data preparation, the simulation run and the analysis of the simulation output files, little file-sharing between users is done.

## 2.3. Where are the Bytes?

This time we looked at a snapshot of the size distribution of all files in the parallel file system scratch space on the TACC TeraGrid cluster. TACC has a policy of purging any files in its scratch file system that has not been accessed for more then ten days. We can therefore assume that the files were actively accessed by running jobs or users. Again, the directories where these files were examined belonged to TeraGrid-wide users, so our conclusions represent one data point for the TeraGrid as well.

**Table 1. Cumulative file size distribution for the parallel file system scratch space on the TACC TeraGrid cluster.**

| Size | Files | Frequency | Total gigabytes | Frequency |
|---|---|---|---|---|
| > 500M | 130 | 0.09% | 302.471 | 35% |
| > 400M | 204 | 0.14% | 335.945 | 38.87% |
| > 300M | 271 | 0.19% | 359.140 | 41.55% |
| > 200M | 1413 | 0.99% | 623.137 | 70.09% |
| > 100M | 2523 | 1.76% | 779.611 | 90.19% |
| > 1M | 12856 | 9% | 851.347 | 98.49% |
| > 0.5M | 16077 | 11.23% | 853.755 | 98.77% |
| > 0.25M | 30962 | 21.62% | 859.584 | 99.45% |
| Total | 143190 | 100% | 864.385 | 100% |

Table 1 summarizes our findings. We note that even though only 9% of the files were greater then 1 megabyte in size, over 98.49% of the bytes are from files in this range. The file bytes represent the used/generated bytes by computational simulation runs in the TeraGrid cluster; hence they represent the majority of the I/O activity for the parallel file system. Out study validates other studies [23] that have also shown this trend towards larger files in scientific computing environments.

## 2.4. Design Assumptions

From our empirical observations we make the following design assumptions. First, file access is more important then file-sharing between users. This is additionally borne out in an internal 2004 TeraGrid survey [24]. If file sharing between users is required, this can be easily achieved by copying files from XUFS to some other space if needed.

Second, file access needs to originate from a user's personal work-space, from a desktop workstation or laptop. A consequence of this is that the file server is now the user's personal system, and we assume this to be unreliable, i.e. disconnects from the server is the norm.

Third, some files should never be copied back to the personal work-space. In our case, we allow the user to specify hints as to which directory any newly generated files are kept local to the client. We call these *localized directories*.

Fourth, client machines can have plenty of disk capacity. User files residing on personal workstations need to be accessed by TeraGrid "client" sites where over 4 PB of combined disk capacity exists.

Fifth, the interface to the local parallel file system should be exposed to the application at the client, enabling the use of any available advanced file system features.

## 3. Detailed Design

### 3.1. Component architecture

The current implementation of XUFS is provided in a shared object, `libxufs.so`, and a user-space file server. XUFS uses the library preloading feature in many UNIX variants to load shared objects into a process image. The XUFS shared object implement functions that interpose the `libc` file system calls, allowing applications to transparently access files and directories across the WAN by redirecting operations to local cached copies.

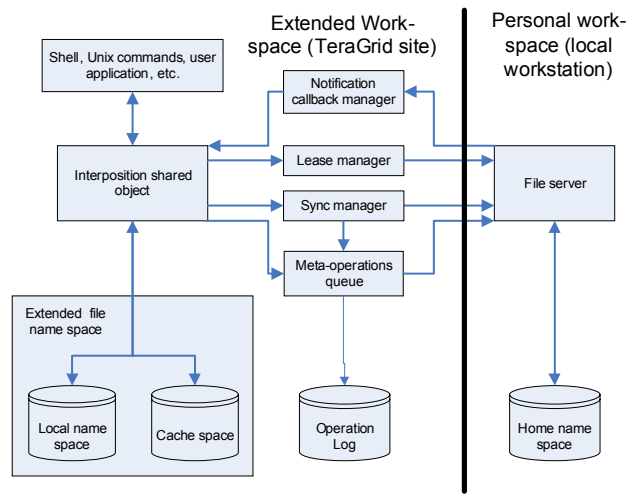

**Figure 1. XUFS component architecture**

The component architecture of XUFS is shown in Figure 1. When an application or user first mounts a remote name space in XUFS, a private *cache space* is created on the client host. At TeraGrid sites, this cache space is expected to be on the parallel file system work partition.

When an `opendir()` is first invoked on a directory that resides in the remote *home name space*, the interposed version of the system call contacts a *sync manager*, downloads the directory entries into the cache space, and redirects all directory operations to this local cache copy. XUFS essentially recreates the entire remote directory in cache space, and stores the directory entry attributes in hidden files alongside the initial empty file entries. Subsequent `stat()` system calls on entries in this directory return the attributes stored in the hidden file associated with each entry. Only when an `open()` is first invoked on a file in this directory will the interposed version of the open system call contact the sync manager again to download the file into the cache space.

System calls that modify a file (or directory) in a XUFS partition return when the local cache copy is updated, and the operation is appended to a persisted *meta-operation queue*. No file (or directory) operation blocks on a remote network call. A `write()` operations is treated differently from other attribute modification operations however, in that the write offsets and contents are stored into an internal *shadow file* and the shadow file flush is only appended to the meta-operations queue on a `close()`. Thus only the aggregated change to the content of a file is sent back to the file server on a close. XUFS therefore implements the *last-close-wins* semantics for files modified in XUFS mounted partitions.

Cache consistency with the home space is maintained by the *notification callback manager*. This component registers with the remote file server, via a TCP connection, for notification of any changes in the home name space. Any change at the home space will result in the invalidation of the cached copy, requiring it to be re-fetched prior to being accessed again. This guarantees the same semantics the user is expected to encounter when the file is similarly manipulated locally.

In the case of a client crash, we provide a command-line tool for users to sync operations, which were in the meta-operations queue at the time of the crash, with the file server. In the case of a server crash, the file server is restarted with a `crontab` job on recovery, with the client periodically attempting to re-establish the notification callback channel when it notices its termination.

File locking operations, except for files in *localized directories*, are forwarded to the file server through the *lease manager*. The lease manager is responsible for periodically renewing the lease on a remote lock to prevent orphaned locks. Files in a localized directory can use the locking mechanisms provided by the cache space file system.

### 3.2. Security

XUFS provides multiple means to access its distributed file system functionality. However, we provide an OpenSSH client, USSH, as one default mechanism. With USSH, users can login from their personal system into a site on the TeraGrid and access personal files from within this login session by mounting entire directory trees from their personal work-space. USSH provides XUFS with a framework for authenticating all connections between client and server.

When USSH is used to log into a remote site, the command generates a short-lived secret `<key,phrase>` pair, starts a personal XUFS file server, authenticates with the remote site using standard SSH mechanisms, and starts a remote login shell. USSH then preloads the XUFS shared object in the remote login shell, and sets the `<key, phrase>` pair in the environment for use by the preloaded shared object. Subsequent TCP connections between the client and file server is then authenticated with the `<key, phrase>` pair using an encrypted challenge string. Communication encryption can be further configured by enabling the use of port-forwarding through a SSH tunnel. This option can also be used for tunneling through local firewalls if needed.

### 3.3. Striped Transfers and Parallel Pre-Fetches

For large latency wide-area networks, the caching of entire files can take a non-negligible amount of time. The situation is further exacerbated by the fact that there are a significant number of very large files in use in our environment. In order to alleviate this potential usability issue, we take advantage of the large network bandwidth available on the TeraGrid.

All data transfers in XUFS over 64 Kbytes are striped across multiple TCP connections. XUFS uses up to 12 stripes with a minimum 64 kilobytes block size each when performing file caching transfers. XUFS also tries to maximize the use of the network bandwidth for caching smaller files by spawning multiple (12 by default) parallel threads for pre-fetching files smaller then 64 kilobytes in size. It does this every time the user or application first changes into a XUFS mounted directory.

## 4. Performance

We perform three experiments to examine the micro and macro behavior of XUFS, compared to GPFS-WAN, on the TeraGrid WAN. The experiments measure the comparative performance of the two file systems using 1) the IOzone micro-benchmarks [29], 2) the build time of a moderate size source code tree, and 3) the turnaround times for accessing a large file located across the WAN.

For the GPFS-WAN scenarios, the benchmarks were all run at NCSA. For the XUFS scenarios, configured with no encryption, we import a directory from the SDSC GPFS scratch space into the NCSA GPFS scratch space, and we ran the benchmarks in the imported directory. Thus in both scenarios, the authoritative versions of the files were always at SDSC (the GPFS-WAN file servers are at SDSC). The network bandwidth between SDSC and NCSA is 30 Gbps.

### 4.1. Micro-Benchmark

The IOzone benchmark measures the throughput of a variety of file operations on a file system. In this experiment we use the benchmark to measure the `read` and `write` performance.

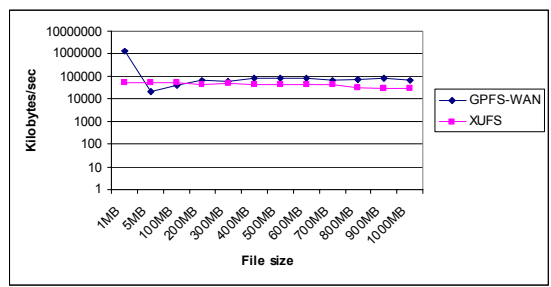

**Figure 2. Write performance of the WAN file systems**

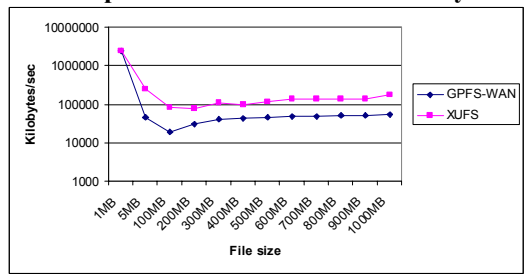

**Figure 3. Read performance of the WAN file systems**

We ran the benchmark for a range of file sizes from 1 MB to 1 GB, and we also included the time of the `close`

operation in all our measurements to include the cost of cache flushes. The throughput for the write and read performance is shown in Figure 2 and Figure 3.

XUFS demonstrates better performance then GPFS-WAN in the read case for files larger then 1 MB. As we have seen, the majority of bytes on the TeraGrid are in files greater then 1 MB, so this is an important benefit. XUFS does well because it directly accesses files from the local cache file system.

Also, XUFS performance is generally comparable to GPFS-WAN for the write case. However, GPFS-WAN demonstrate far better write performance than XUFS for the 1 MB file. This is probably demonstrates the benefit of memory caching in GPFS.

### 4.2. Source Code Build Times

In this experiment we built a source code tree, containing 24 files of approximately 12000 lines of C source code distributed over 5 sub-directories. A majority of the files in this scenario were less then 64 KB in size. In our measurements we include the time to change to the source code tree directory and perform a clean "make". This experiment measures the amortized overhead of an expected common use case on the WAN file systems.

The timings of consecutive runs are shown in Figure 4. Surprisingly XUFS mostly out performs GPFS-WAN in this benchmark. We speculate this is due to our aggressive parallel file pre-fetching strategy.

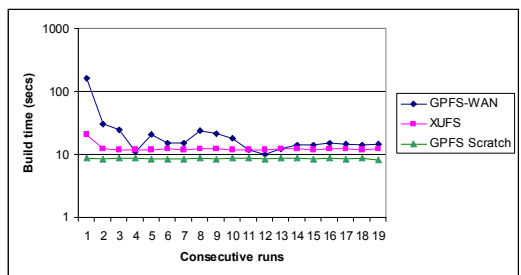

**Figure 4. Build times on the WAN file systems and the local GPFS partition.**

### 4.3. Large File Access

In this experiment we measured the time of a shell operation on a 1 GB file on the WAN file systems. The command used was "wc -l": the command opens an input file, counts the number of new line characters in that file, and prints this count. The timings of consecutive runs of the command are shown in Figure 5.

The results show the command takes approximately 60 seconds to complete in the first instance when run in XUFS. This delay is due to the system copying the file into its cache space for the first invocation of open(). However, subsequent invocation of the command performs considerably better because XUFS is redirecting all file access to the local cache file. GPFS-WAN show a consistent access time of 33 seconds in all 5 runs of the experiment.

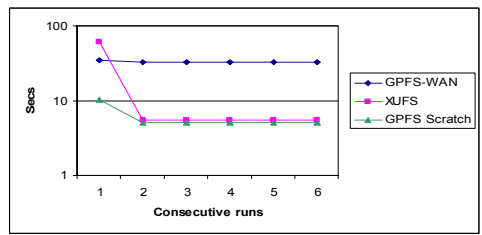

**Figure 5. Access timings for a 1GB file on the WAN file systems and the local GPFS partition.**

We also compared the XUFS timings, with the timing from manually copying the 1GB file from SDSC to NCSA using the copy commands TGCP (a GridFTP client) and SCP. The timings are summarized in Table 2. The results show TGCP as having only a slight advantage over XUFS in terms of the turnaround time of accessing the large file. The poor performance of SCP is probably due to its use of encryption, and its lack of TCP stripping.

**Table 2. Mean time of "wc -l" on a 1GB file in XUFS, compared to copying the file across the WAN using the TGCP and SCP copy commands.**

| XUFS (secs) | TGCP (secs) | SCP (secs) |
| --- | --- | --- |
| 57 | 49 | 2100 |

### 5. Related Work

The XUFS approach of using a shared object to interpose a global name space for file systems located across a WAN is similar to the approach expounded by the Jade file system [22]. Like Jade, XUFS allows private name spaces to be created, and aggressively caches remote files to hide access latency on the WAN. However, unlike Jade, XUFS employs a different cache consistency protocol, and uses striped block transfers and parallel file pre-fetching to better utilize large bandwidth networks.

XUFS caches entire files on disk instead of using smaller memory block caches. Most distributed file systems cache memory blocks for fast repeat access [5][6][15][17][18]. A few past systems have also used entire file caches; e.g. Cedar [28], AFS-1 [14], and Coda [16]. Our primary reason for caching entire files is because of our assumption that the server is unreliable. Computation jobs need reliable access to input data files, and having the entire file in cache allows XUFS to provide access to files even during temporary server or network outages.

XUFS reliance on a callback notification mechanism for its cache consistency protocol is similar to the approach used by AFS-2 and AFS-3 [30]. Cached copies are always assumed to be up-to-date unless otherwise notified by the remote server. This protocol is different from the cache consistency protocol in NFS and Jade where clients are responsible for checking content versions with the remote server on every file open(). This is also different from the token-based cache consistency protocols used in GPFS [6], Decorum [15] and Locus [27]. Token-based consistency protocols are particularly efficient for multiple-process write-sharing scenarios. However, XUFS allows the high performance writing-sharing features in a parallel file system, configured as the cache space, to be made available to an application. This satisfies the overwhelming majority of cases where an application is scheduled to run on multiple CPUs at one site. Exposing the local file system features in the distributed file system interface is also supported by the Decorum file system [15] and extensions to the NFSv4 [25] protocol.

XUFS share similar motivation with other recent work on wide-area network file systems, such as Shark [33]. However, XUFS differs with Shark in its fundamental design goals and detailed implementation. Shark uses multiple cooperating replica proxies for optimizing network bandwidth utilization and alleviating the burden of serving many concurrent client requests on the central file server. XUFS uses stripped file transfers to maximize network bandwidth utilization, and instantiates a private user-space file server for each user.

Finally, the semantics of our USSH access client is similar in many respects to the `import` command in Plan 9 [13], and to the remote execution shell *Rex* [32]. However, USSH differs from these other commands in the underlying protocols used to access files and directories: 9P in the case of `import`, and SFS in the case of *Rex*.

## 6. Conclusion

XUFS is motivated by real user requirements; hence we expect the system to evolve over time based on feedback from scientists using it while engaged in distributed computational research. However, future work on XUFS will also investigate integration with alternative interposition mechanisms such as SFS [31], FUSE [35] and Windows *Detours* [34]. These will broaden the range of XUFS supported platforms and usage scenarios, eliciting a wider community of users to instigate its evolution.

## Acknowledgement

We would like to thank the anonymous reviewers for their helpful and constructive comments. In particular we would like to thank our shepherd, Mike Dahlin, for diligently reviewing and providing valuable feedback to the final manuscript.